\documentclass[twoside]{ae100prg}
\usepackage[breaklinks]{hyperref}
\usepackage{booktabs}
\usepackage{graphicx,amsmath,amssymb}

\begin{document}
\title{Non-Linear Effects in Non-Kerr spacetimes}

\author{Georgios Lukes-Gerakopoulos$^1$, George Contopoulos$^2$,
Theocharis A. Apostolatos$^{3}$}

 \address{$^1$ Theoretical Physics Institute, University of Jena,
 \\ 07743 Jena, Germany}
 \address{$^2$  Research Center for Astronomy, Academy of  Athens,
  Soranou Efesiou 4, GR-11527 Athens, Greece}
 \address{$^3$ Section of Astrophysics, Astronomy, and Mechanics,
 Department of Physics, University of Athens, Panepistimiopolis Zografos GR15783,
 Athens, Greece}

\email{gglukes@gmail.com}

\begin{abstract}
 There is a chance that the spacetime around massive compact objects which are
 expected to be black holes is not described by the Kerr metric, but by a metric
 which can be considered as a perturbation of the Kerr metric. These non-Kerr
 spacetimes are also known as bumpy black hole spacetimes. We expect that, if some
 kind of a bumpy black hole exists, the spacetime around it should possess some
 features which will make the divergence from a Kerr spacetime detectable. One of
 the differences is that these non-Kerr spacetimes do not posses all the symmetries
 needed to make them integrable. We discuss how we can take advantage of this fact
 by examining EMRIs into the Manko-Novikov spacetime. 
\end{abstract}

\section{Introduction}

We expect that a star which at the end of its life becomes a compact object with
mass greater than three solar masses is a Kerr black hole. However, this
anticipation should be somehow tested by observations.

One way to test the Kerr hypothesis is to study the gravitational wave signal
produced by an inspiraling relatively light compact object (e.g., stellar) into
the spacetime background of a supermassive compact object. This kind of motion
is called Extreme Mass Ratio Inspiral (EMRI). Such binary systems should exist
in the center of galaxies which we believe are occupied by supermassive black
holes ($10^5$ to $10^9$ solar masses). In EMRIs the lighter object basically
traces the background spacetime by following approximately geodesic orbits. Ryan
in \cite{Ryan95,Ryan97} showed that we could extract the multipole moments of the
background from the gravitational wave signal, and Collins and Hughes
\cite{Collins04} produced a perturbed Schwarzschild black hole spacetime, which
they called ``bumpy'' black hole spacetime, in order to perform the first tests
of the Kerr hypothesis. Since then several such tests have been proposed, see e.g.,
\cite{eLISA,Bambi11,Johan12} and references therein.

The bumpy black hole spacetimes are axisymmetric and stationary, but in general
lack a Carter-like constant \cite{Carter68}. It has been shown that even in an 
axisymmetric stationary Newtonian potential, a higher order Killing tensor
connected to a Carter-like constant cannot be found \cite{Markakis}, contrary to
conjectures that were initially postulated \cite{Brink}. The lack of a Carter-like
constant implies that the bumpy black hole systems are non-integrable, which in
turn suggests that non-linear effects like chaos should be present. In a series of
publications \cite{ALGC09,LGAC10,CLGA11} we have studied what implications these
non-linear effects will bring to a gravitational wave signal coming from an EMRI
into a non-Kerr spacetime background. In the present article we present briefly
these findings. 

The article is organized as follows. Section \ref{sec:ThEl} introduces some basic 
theoretical elements about a bumpy black hole spacetime and the geodesic motion
in such spacetime. Section \ref{sec:NonEff} discusses the non-integrability
imprints of the non-Kerr background in gravitational wave signals. Our conclusions
are given in Section \ref{sec:Con}.

\section{Theoretical elements} \label{sec:ThEl}

\subsection{The Manko-Novikov spacetime} \label{subsec:MNspa}

The bumpy black hole spacetime we used in \cite{ALGC09,LGAC10,CLGA11} is a
spacetime which belongs to the so-called Manko-Novikov (MN) metric family
\cite{ManNov92}. Manko and Novikov found an exact vacuum solution of Einstein's
equations which describes a stationary, axisymmetric, and asymptotically flat
spacetime with arbitrary mass-multipole moments \cite{ManNov92}. The MN metric
subclass we used was introduced in \cite{Gair08} and deviates from the Kerr at all
moments higher or equal to the quadrupole one. The new spacetime is characterized
by one more parameter $q$ than the ones describing a Kerr metric. Namely, the
quantity $q$ measures how much the MN quadrupole moment $Q$ departs from the Kerr
quadrupole moment $Q_{Kerr}=-S^2/M$ (that is $q=(Q_{Kerr}-Q)/M^3$), where $M$ and
$S$ are the mass and the spin of a Kerr black hole respectively. If $q=0$ the MN 
solution becomes exactly a Kerr solution. The line element of the MN metric in the
Weyl-Papapetrou cylindrical coordinates $(t, \rho, \varphi, z)$ is 
 \begin{equation} \label{eq:MNmetric}
  ds^2=-f(dt-\omega d\varphi)^2 + f^{-1} [e^{2\gamma} (d\rho^2 + dz^2)
       +\rho^2 d\varphi^2],
 \end{equation}
where $f,~\omega,~\gamma$ are considered functions of the prolate spheroidal
coordinates $v, w$, while the coordinates $\rho,z$ can be expressed as functions
of $v, w$ as well. Namely,
%
%\begin{equation}\label{eq:trans}
 $\rho=k \sqrt{(v^2-1)(1-w^2)}$, $z=k v w$,
%\end{equation}
%
%and
%
%\begin{subequations}
%\begin{eqnarray}
% f &=& e^{2 \psi}\frac{A}{B}, \label{ffunc} \\
% \omega &=& 2 k e^{-2 \psi}\frac{C}{A}-4 k \frac{\alpha}{1-\alpha^2}, \\
% e^{2 \gamma} &=& e^{2 \gamma^\prime}\frac{A}{(v^2-1)(1-\alpha^2)^2},
% \label{fexpgam} \\
% A &=& (v^2-1)(1+a~b)^2-(1-w^2)(b-a)^2,\label{fA} \\
% B &=& [(v+1)+(v-1)a~b]^2+[(1+w)a+(1-w)b]^2,\label{fB} \\
% C &=& (v^2-1)(1+a~b)[(b-a)-w(a+b)] \nonumber \\
%   &&+ (1-w^2)(b-a)[(1+a~b)+v(1-a~b)], \\
% \psi &=& \beta \frac{P_2}{R^3}, \label{fC}\\
% \gamma^\prime &=& \ln{\sqrt{\frac{v^2-1}{v^2-w^2}}}+\frac{3\beta^2}{2 R^6}
% (P_3^2-P_2^2) \nonumber \\ &+& \beta \left(-2+\displaystyle{\sum_{\ell=0}^2}
% \frac{v-w+(-1)^{2-\ell}(v+w)}{R^{\ell+1}}P_\ell\right), \label{fgampr}\\
% a &=& -\alpha \exp {\left[-2\beta\left(-1+\displaystyle{\sum_{\ell=0}^2}
% \frac{(v-w)P_\ell}{R^{\ell+1}}\right)\right]}, \label{fa}\\
% b &=& \alpha \exp {\left[2\beta\left(1+\displaystyle{\sum_{\ell=0}^2}
% \frac{(-1)^{3-\ell}(v+w)P_\ell}{R^{\ell+1}}\right)\right]}, \label{fb}\\
% R      &=& \sqrt{v^2+w^2-1}, \label{fR}\\
% P_\ell &=& P_\ell (\frac{v~w}{R}). \label{fLegA}
%\end{eqnarray}
%\end{subequations}
%
%Here $P_\ell(\zeta)$ is the Legendre polynomial of order $l$
%
%\begin{equation} \label{fLeg}
% P_\ell(\zeta)=\frac{1}{2^\ell \ell!}
%\left(\frac{d}{d\zeta}\right)^\ell(\zeta^2-1)^\ell,
%\end{equation}
%
%while the parameters $k,\alpha,\beta$ are related to the mass $M$, the spin $S$,
%and the quadrupole deviation $q$ through the expressions
%
where
%
%\begin{equation}
%\begin{array}{r}
 $k=M\frac{1-\alpha^2}{1+\alpha^2}$,
%\end{array}
%\begin{array}{c}
 $\alpha=\frac{-1+\sqrt{1-\chi^2}}{\chi}$,
%\end{array}
%\begin{array}{l}
% \beta=q \left( \frac{1+\alpha^2}{1-\alpha^2} \right)^3.
%\end{array}
%\label{eq:freepar}
%\end{equation}
%
while $\chi$ is the dimensionless spin parameter $\chi=S/M^2$. The exact formulae
of $f,~\omega,~\gamma$ are lengthy, and can be found in \cite{Gair08,LGAC10}.  

\subsection{Geodesic motion in the Manko-Novikov spacetime} \label{subsec:MNgeo}  

The geodesic orbits of a test particle of mass $\mu$ are described as equations
of motion of the Lagrangian
%
%\begin{equation}
$L=\frac{1}{2}~\mu~g_{\mu\nu}~ \dot{x}^{\mu} \dot{x}^{\nu}$,
%\label{LagDef}
%\end{equation}
%
where the dots denote derivatives with respect to the proper time. The MN metric
has two integrals of motion, namely the energy (per unit mass)
\begin{equation} \label{eq:EnCon}
E=-\frac{\partial L}{\partial \dot{t}}/\mu=
f (\dot{t} - \omega~ \dot{\varphi}),
\end{equation}
and the z-component of the angular momentum (per unit mass)
\begin{equation} \label{eq:AnMomCon}
L_z =\frac{\partial L}{\partial \dot{\varphi}}/\mu=
f \omega (\dot{t} - \omega~ \dot{\varphi})+ f^{-1} \rho^2 \dot{\varphi},
\end{equation}
The Kerr metric has one more integral of motion, the so-called Carter constant
\cite{Carter68}, thus it is an integrable system. However, the MN model lacks in
general (as long as $q \neq 0$) such constant, which means that MN is a
non-integrable system, and therefore chaos should appear.

We can reduce the four degrees of freedom of the MN system to two, by using the two
integrals of motion $E$, and $L_z$, and thus, restrict the motion on the meridian
plane ($\rho,~z$).  By rewriting the metric (\ref{eq:MNmetric}) we see that the
motion on the meridian plane satisfies the relation
\begin{equation} \label{eq:Veff}
\frac{1}{2} (\dot{\rho}^2 + \dot{z}^{2}) + V_{eff} (\rho, z)=0,
\end{equation}
where the effective potential $V_{eff} (\rho, z)$ depends on the parameters $q$,
$\chi$, $E$, and $L_z$. $V_{eff}\le 0$ for all possible orbits. On the boundary
$V_{eff}=0$ the velocity vanishes: $\dot{\rho}=\dot{z}=0$; this is the so called
curve of zero velocity (CZV).

In the Kerr spacetime case ($q=0$) every non-plunging geodesic orbit confined by
the CZV lies on a two dimensional torus in the phase space. On such a torus each
orbit is described by two characteristic frequencies $\omega_1,~\omega_2$. If the
ratio of these frequencies $\nu_\theta$ is an irrational number, the motion is 
quasiperiodic, and the corresponding torus is covered densely by the orbit. If the
ratio is a rational number, the motion is periodic, and the corresponding torus is
called resonant. A resonant torus is covered by an infinite number of periodic
orbits, all having the same frequency ratio $\nu_\theta$. 

By setting $q\ne 0$ we perturb the integrable system (Kerr), and the transition to
the non-integrable system (MN) is described basically by two theorems: the KAM
theorem, and the Poincar\'{e}-Birkhoff theorem. The first theorem states that after
the perturbation most of the non-resonant tori will survive deformed. These
surviving tori are called KAM tori. The second theorem implies that from a resonant
torus only a finite even number of periodic orbits will survive; half of them will
be stable and the other half unstable. Around the stable orbits small islands of
stability are formed, while the asymptotic manifolds emanating from the unstable
periodic orbits fill a region of chaotic orbits. The above formation is known as
a Birkhoff chain. 

One way to study the aforementioned different structures in a non-integrable system
of two degrees of freedom is to take a section through the foliation of the tori.
Such section is known as a surface of section, or as a Poincar\'{e} section.
Another way is provided by the frequency ratio $\nu_\theta$, by which we can detect
the different types of orbits and it is known as the rotation number (e.g., 
\cite{Contop02}). If $\nu_\theta$ corresponds to an irrational number, we have a
KAM curve; if it corresponds to a rational number, we are on a Birkhoff chain of
islands of stability; if the value of $\nu_\theta$ is indefinite, and does not
correspond to a particular number, then the orbit is chaotic. 

\begin{figure}
  \begin{center}
  \includegraphics[width=0.8\textwidth]{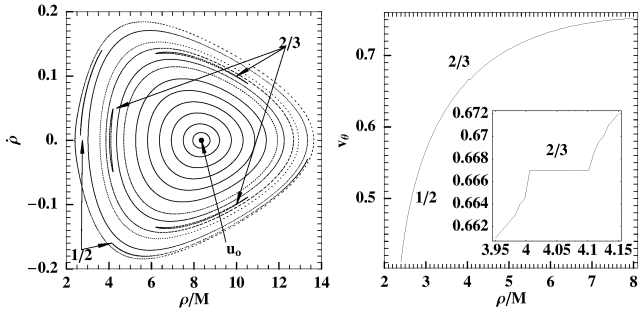}
  \end{center}
  \caption{The left panel shows a part of the surface of section on the plane
   $(\rho,\dot{\rho})$ focusing on the main island of stability, where
   $\mathbf{u}_0$ indicates the center of the main island. The right panel shows
   the rotation number along the line $\dot{\rho}=0$ (starting from $\mathbf{u}_0$
   and moving leftwards) on the surface of section shown in the left panel.
   Embedded in the right panel is a detail of the rotation curve around the 
   $2/3$-resonance. The parameter used are $E=0.95$, $Lz=3~M$, $\chi=0.9$, $q=0.95$.}
\label{FigExtReg} 
\end{figure}

\section{Non-integrability imprints on the gravitational wave signal} 
\label{sec:NonEff}  

\subsection{The plateau effect of the resonances}

One possible imprint of the non-integrability of a bumpy black hole spacetime on
the corresponding gravitational wave signal is the effect of the resonances. The
left panel of Fig.~\ref{FigExtReg} shows a part of the surface of section $z=0$
($z>0$) of a MN spacetime for the parameter set $E=0.95$, $Lz=3~M$, $\chi=0.9$,
$q=0.95$.The first impression one might get is that the surface of section
indicates an integrable system, because no straightforward signs of chaos are
prominent. However, the islands of stability (left panel of Fig. \ref{FigExtReg})
imply the existence of Birkhoff chains, which in turn indicate that chaos is also
present.

If we take initial conditions along a straight line in the phase space, like the
$\dot{\rho}=0$ line starting from the center $\mathbf{u}_0$ of the main island of
stability shown in the left panel of Fig. \ref{FigExtReg}, and evaluate the
rotation number for each of these initial conditions, then we get a rotation curve
(right panel of Fig. \ref{FigExtReg}). This curve seems to be smooth and strictly
monotonic (in a Kerr spacetime this is the case), however a more detailed look
reveals that this is not exactly true. At the resonances, plateaus appear. For
instance in the embedded plot of the right panel of Fig. \ref{FigExtReg}, we can
see a plateau at the $2/3$ resonance. This happens because all the orbits belonging
to the same chain of islands of stability share the same rotation number, i.e., the 
same frequency ratio. Such plateaus do not appear in the case of a Kerr metric.

However, geodesic orbits are simply an approximation of real EMRI orbits. A more
realistic approximation demands the inclusion of the radiation reaction. Since
there are no reliable computations describing the radiation reaction in a bumpy
black hole spacetime, we used the same trick with the authors of \cite{Gair08}.
Namely, we used the hybrid approximative method \cite{Gair06} (eqs. (44), (45) in
\cite{Gair06}), where we added by hand the anomalous quadrupole moment $q$ to the
$\chi^2$ terms. Furthermore, we assumed a constant rate of energy and angular
momentum loss due to gravitation radiation.

\begin{figure}[htp]
  \begin{center}
  \includegraphics[width=0.4\textwidth]{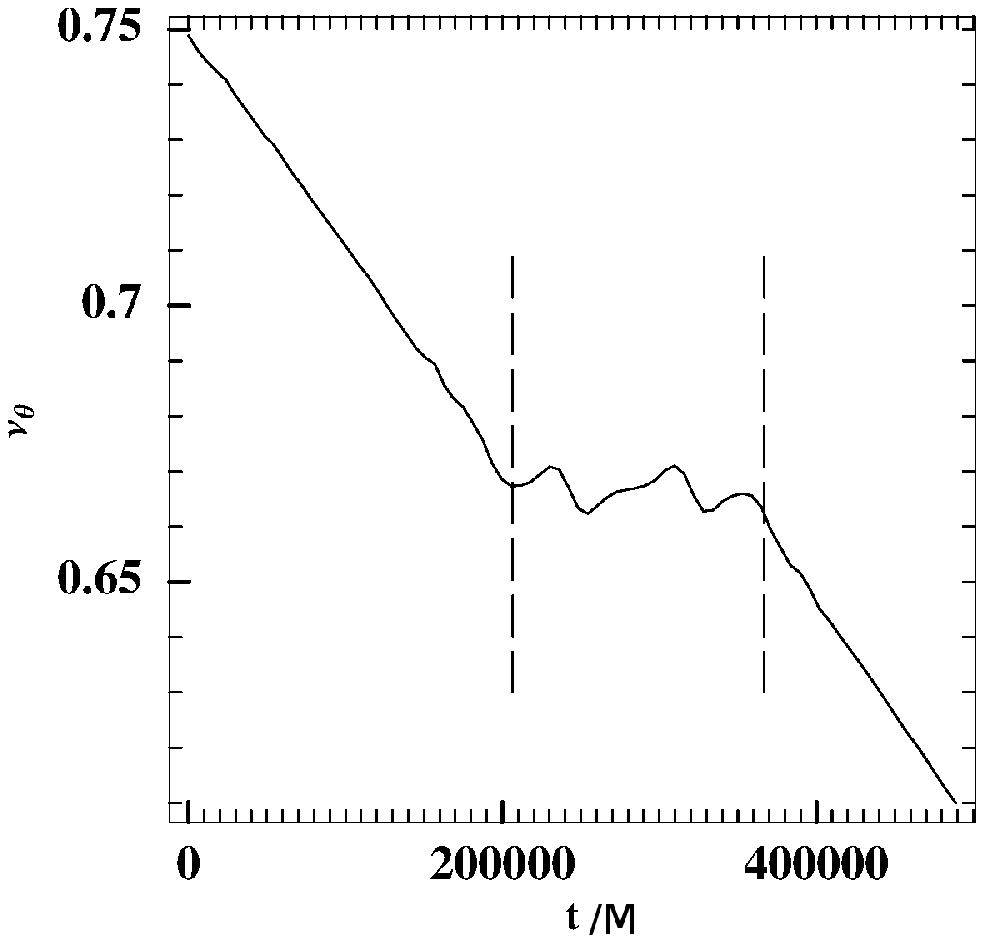}
  \includegraphics[width=0.4\textwidth]{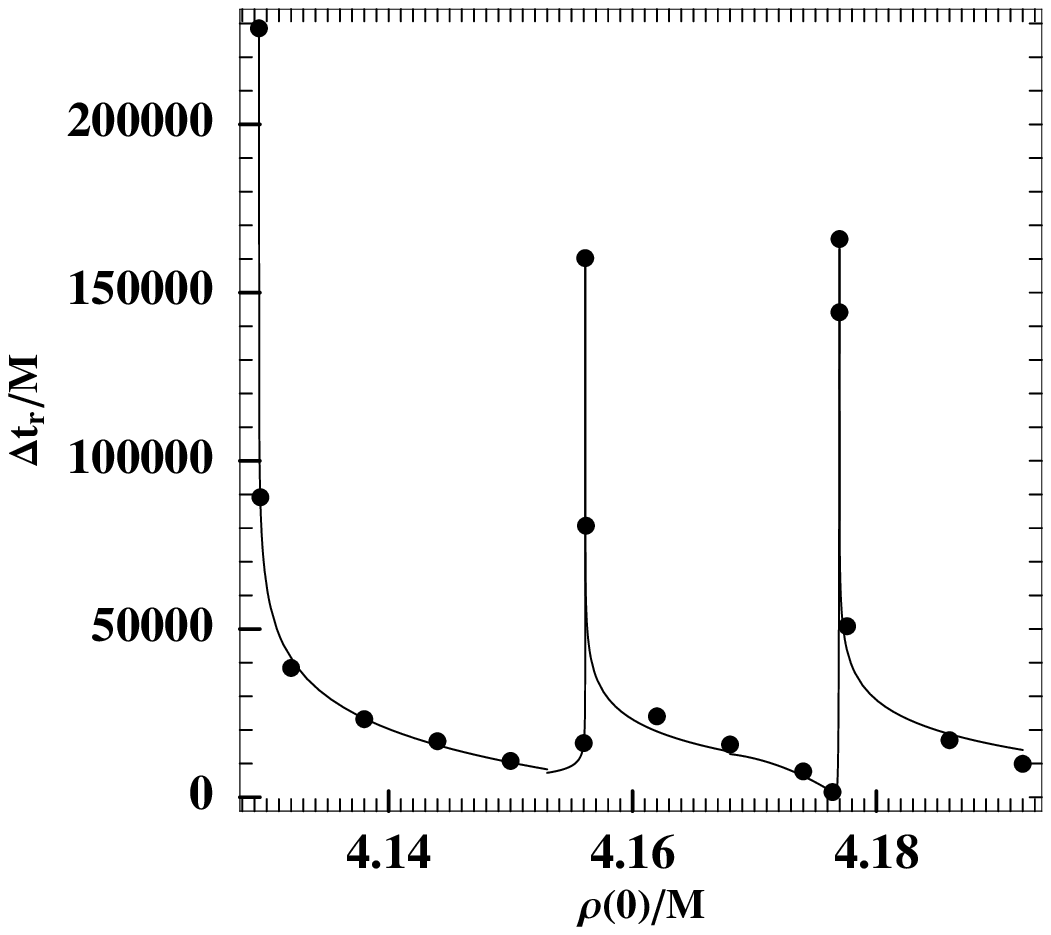}
  \end{center}
  \caption{The left panel shows the evolution of the ratio $\nu_\theta$ as
   a function of the coordinate time $t$ for a non-geodesic orbit. The vertical
   dashed lines demarcate the time intervals that the non-geodesic orbit spends in
   the interior of the $2/3$-resonance. The right panel shows the time $\Delta t_r$
   needed by non-geodesic orbits to cross the chain of islands belonging to the
   $2/3$-resonance as a function of their initial conditions $\rho(0)$ along the
   line $\dot{\rho}=0,~z=0$.  The parameters used are $\mu/M=8 \times 10^{-5}$,
   $q=0.95$, $\chi=0.9$, $E(0)=0.95$, $L_z(0)=3~M$, where $E(0),~L_z(0)$ are the
   initial values of $E$ and $L_z$ respectively.}
 \label{FigNGTi}
\end{figure}

We applied the aforementioned scheme for a mass ratio $\mu/M=8 \times 10^{-5}$,
to evolve initial conditions near the $2/3$ resonance, and found that the plateau
also appears when we calculate the rotation number as a function of the coordinate
time (left panel of Fig. \ref{FigNGTi}). This phenomenon was tested for several
initial conditions near the $2/3$ resonance and for each of them we estimated the
time $\Delta t_r$ that the inspiraling non-geodesic orbit stayed in the resonance
(right panel of Fig. \ref{FigNGTi}). The mean time of these plateaus is
approximately $5 \times 10^4 M$, which corresponds roughly to a week for a
supermassive compact object of the size of the one lying at the center of the Milky
Way.   

\subsection{The Beacon effect of stickiness}

\begin{figure}[htp]
  \begin{center}
  \includegraphics[width=0.8\textwidth]{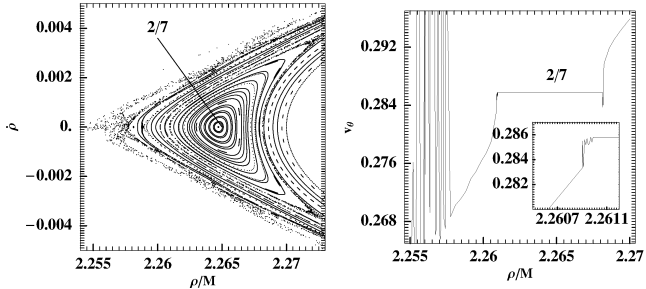}
  \end{center}
  \caption{ The left panel shows a detail of the surface of section near an island
   $2/7$. The right panel shows the rotation curve along the $\dot{\rho}=0$ line of
   the surface of section presented in the left panel. Embedded in the right panel
   is the irregular variation of the rotation number just outside the left side of
   the $2/7$-plateau. This irregular behavior is due to the chaotic layer
   surrounding the corresponding island. The parameters used are $E=0.95$,
   $Lz=2.995~M$, $\chi=0.9$,  $q=0.95$.}
\label{FigNeck}
\end{figure}

If we focus more on the chaotic aspect of the Birkhoff chains, another effect could
be detected in gravitational waves coming from an EMRI in a bumpy black hole
spacetime background. This effect is connected with the phenomenon of stickiness
\cite{Contop02}. The stickiness phenomenon concerns chaotic orbits which for
various reasons stick for a long time interval in a region, close to  regular
orbits. Therefore, their behavior in the frequency spectrum might resemble that of
the regular orbits they are close to, before they depart from that region.

In the left panel of Fig. \ref{FigNeck} we see a detail of the surface of section
near the resonance $2/7$. The stickiness appears in the region where chaotic orbits
(scattered points on the surface of section) are confined by regular orbits. Even
though their true character is detected by the rotation number, since $\nu_\theta$
varies widely in the corresponding regions (right panel of Fig. \ref{FigNeck}), the
phenomenology might be more complicated in the frequency spectrum. Namely, while a 
chaotic orbit stays near a regular orbit we might get a signal, i.e., distinct
characteristic frequencies; when the orbit moves to a more prominent chaotic layer 
the frequency peaks in the signal will dissolve leaving only noise instead of a
signal; later on when the orbit will return near a regular orbit the signal shall
reappear, and so on. This effect resembles a beacon, where the signal appears and
disappears.
       
\section{Conclusions} \label{sec:Con}

The resonance and the stickiness effect are generic characteristics of the geodesic
motion in any non-integrable Hamiltonian system describing a stationary and
axisymmetric spacetime background like that of an axially symmetric perturbation of
the Kerr spacetime. Therefore, they should be in principle detectable in the
gravitational wave signal coming from an EMRI into a non-Kerr metric.

\vspace{1cm}
{\bf Acknowledgements}: G. L-G is supported by the DFG grant SFB/Transregio 7.

\section*{References}
\bibliography{NonKerr}

\end{document}